# 基于区块链的电网大数据数字资产管理架构

张俊 [1,2,3]　王飞跃 [1,3]

**摘要**：由于受到科学技术和企业架构的约束，中国电网企业多年来对电网大数据的数字资产管理研究开发收效甚微。区块链的去中心化、去信任、信息难以篡改等技术特性，为电网大数据的数字资产管理以及进一步推动电网大数据的应用，提供了一条崭新而极有前景的技术路径。本文首先介绍区块链技术、电网大数据以及数字资产管理理论研究的发展现状，然后总结当前电网企业在数字资产管理过程中存在的主要问题，并针对这些问题制定相应的基于区块链技术的解决方案和用于执行此方案的管理架构，并对区块链技术在电网数字资产管理方面的未来发展方向进行了分析和总结。
**关键词**：区块链，电网大数据，数字资产管理

## 1 引言

### 1.1 区块链技术

区块链，是一种全网节点共同维护用于存储历史交易记录或数据信息的分布式共享超级账本，其通过采用分布式共识机制、非对称加密算法、区块链式存储等技术实现了去中心化（全中心化）、去信任化（全信任化）、数据信息难以篡改等功能。

自 2008 年中本聪提出去中心化的点对点交易体系以来，加密货币体系比特币引起

1. 中国科学院自动化研究所复杂系统管理与控制国家重点实验室 北京 100190　2.武汉大学电气工程学院 武汉 430072　3.青岛智能产业技术研究院青岛 266071
1. The State Key Laboratory of Management and Control for Complex System, Institute of Automation, Chinese Academy of Sciences, Beijing 100190　2. School of Electrical Engineering, Wuhan University, Wuhan 430072　3. Qingdao Academy of Intelligent Industries, Qingdao 266071

了各界研究人员的广泛关注，其底层技术，区块链，也逐渐走入人们的视野[1]。中本聪通过区块链技术将全球人与人之间的信任问题转移到了对分布式共识机制—工作量证明机制上，实现了虚拟货币比特币在全球范围内的安全流通，充分诠释了"信任是最有效的流通货币"这一句话。如今，世界上很多知名电子商务平台已经相继承认比特币作为可支付使用货币，德国政府也在2013年承认了其合法的货币地位。比特币的成功带动了更多的虚拟加密货币的涌现，也激发了研究人员对于区块链在货币的流通与交易中应用的研究兴趣，目前人们研究最多的区块链理论也是基于为虚拟货币服务的区块链技术。

区块链技术虽然是作为加密货币比特币的底层技术引起研究人员的广泛关注的，但其应用领域并非仅限于虚拟货币的交易。通过利用区块链技术去信任的特点与可灵活编程的智能合约技术相结合[2]，可将区块链技术的应用领域从单纯的虚拟货币交易扩展到与货币交易相类似的其他金融交易中。当交易预设条件达成时，部署在区块链上的智能合约将强制性自动执行合约内容，有效避免了第三方的介入，能够广泛应用于证券、众筹、借贷等对信用要求高的金融交易中，也是目前区块链技术应用研究的一个重要领域。

目前，区块链技术除了在金融交易领域中的到应用外，在其他很多领域都获得了应用，具有广阔的应用前景，其应用范围包括物联网身份认证、食品药品监管、数据资产管理等，一切需要去中心化、去信任、第三方监管的问题，均能通过区块链技术得到解决。自此，区块链技术的应用领域完成了从金融领域到其他领域的拓展。

## 1.2 电网大数据

电网大数据是指产生于智能电网系统中各个环节中的海量的多源异构数据，其具有"4V"特征，即规模大、类型多、价值稀疏以及变化快[3]。

近年来，为了应对全球能源问题，世界各国全面开展智能电网的研究工作[4]，建立覆盖电力系统整个生产过程的智能电网是电力系统的首要任务。由于电网大数据是对智能电网的安全稳定运行起支撑作用的基础，当前，电网大数据已成为学术界与电力行业共同关注的研究课题，并在多处电力生产环节中得到应用，具有十分广阔的应用前景。对电网大数据的研究主要分为两个方面：其一，由于电网大数据来源于智能电网中海量的智能电表、传感器、电气装备等设备，具有体量大、类型复杂、速度快等特点，如何对其进行高效的管理是对电网大数据的研究过程中的一大难题；其二，电网大数据具有很高的利用价值，通过对电网数据的分析，可以在电网的运行、运营、供给侧、传输侧、需求侧各个方面和层面上利用，例如可以预测光伏与风力发电的输出功率，提高电网对光伏和风电的接纳能力，优化电网运行方式和潮流，也可以学习电力用户用电规律，合理设置电力需求响应系统，甚至可优化电力企业内部管理结构[5]等等。

2013年，中国电机工程学会电力信息化专委会编制并发布了《中国电力大数据发展白皮书》，白皮书中详细定义了电网大数据的特征以及发展路线，对电网大数据的研究已成为我国电网发展的迫切需求。自电网大数据白皮书发布以来，电力行业及学术机构

对电网大数据的分析技术进行了大量的研究，然而对电网大数据的应用前提应是对电网大数据高效的采集、存储以及管理，而目前电力系统在对电网大数据的存储与管理上依旧存在许多问题：数据定义不统一、数据存储内容不一致、数据内容质量不高等都给电网大数据在电力系统中的应用带来了实质性的阻碍。

### 1.3 数字资产管理

数据资产是企业或机构在生产、运营、管理过程中累积的对企业或机构具有利用价值的数字化信息和内容，通过对数字资产的组织加工，可以优化企业内容管理架构，促进企业运营模式改革，从而提高企业收益。

简单地将海量的数字资产存储在各种存储介质中并且不采取任何管理措施，企业的数字资产无法体现其自身的任何价值，为了发挥企业数字资产的最大价值，数字资产管理应运而生。数字资产管理是对数字资产的创建、采集、组织、存储、利用和清除过程加以研究并提出的相应方法的统称，其在数字资产中的应用覆盖了数字资产的整个生命周期，包括使用元数据对数据内容进行描述，实现高效的数据检索功能；通过将信息内容备份存储于不同存储介质中，实现数字资产存储的可靠性；以及通过清除和迁移价值衰减的信息，节约数据存储资源等。

根据数字资产管理的特点，但凡满足信息内容产生于不同部门、对信息内容检索频率较高、需要通过数据驱动的企业均可通过数字资产管理来提高企业拥有的数字资产价值，电网企业就是其中之一，而作为电网企业的数字资产，对电网大数据进行高效的数字资产管理便显得十分重要。

## 2 数字资产管理在电网大数据中的应用

### 2.1 传统电网大数据管理弊端

电网大数据被认为是支撑智能电网安全、稳定、可靠运行的基础，对电网大数据的合理开发和应用能够促进电网发展优化改革，提高电网运行效益。而对电网大数据的合理应用要求高效的数据管理，方便数据收集、检索和更改，目前也有多个地区建立了电网大数据平台，但均采用的是集中式的管理方法，例如北京、上海、山西建立的大数据平台，以及湖北省的全数据中心等。完整的数字资产管理应覆盖数据的创建、采集、组织、存储、利用、清除各个环节，由于电网大数据的数据源分布范围广且产生速度极快，单纯靠集中式的管理方法对整个流程进行管理难度极大。本文根据以往文献并根据实际经验，简要分析目前电网大数据数字资产管理过程中出现的问题：

（1）数据收集质量问题

由于电网大数据数据源分布范围广，需由各地市公司分散完成数据的采集和上传工作，地市公司间缺乏交流。为确保各地市公司收集到的数据在格式以及定义上得到统一，往往需要中心机构给出数据的标准定义，但由于电网公司层级式的管理特点，中心机构与底层地市公司之间的信息传递不畅，容易出现信息传递缓慢甚至失真的严重情况，致使各地市公司对数据的理解不一致，其收集上来的电网数据也就无法统一，数据难以用于分析应用。另外，由于电网数据收集监管机制松散，没有切实可行的奖惩制度和数据校验方法，数据采集过程中容易出现

工作人员消极怠工，不根据现场勘测实际情况进行数据填报，或者私自伪造项目数据，导致电网数据可信度极低，降低数据可利用价值。

（2）数据分享问题

发挥电网大数据价值的必要环节是对电网大数据的开发利用，而对电网数据的访问和共享是对其进行开发利用的基础，例如，研究机构在与电网企业合作进行项目研究时，需要实际电网大数据对其理论研究进行验证支撑；地市公司在处理自己地区电网出现的问题时，需要借鉴和分析其他地区电网的数据。而由于电网企业层级式的管理架构以及分享过程中的信任问题，研究机构和电网公司在访问和共享电网数据时存在严格的限制，这一过程需要花费大量的人力资源和时间用于权限审查和数据校验，严重影响了对电网大数据的价值挖掘。

（3）数据安全性问题

数字资产是企业重要的私有资产，数字资产的安全性若得不到保障，会给企业带来巨大的损失，电网企业更是如此。电力系统行业作为关系到国家民生的行业，其拥有的数字资产若遭到恶意盗取，不仅会使电网企业遭受损失，还会对电力市场造成不良影响，所以，电网大数据的安全保障是一项不可忽视的工作。当前电网企业采取的安全保护措施主要是对数据进行对称加密存储与传输，然而这种加密方法易于破解，且为降低被破解的概率，需要定时对其进行升级，其成本和效用均不理想。

目前电网企业以及很多学者重视的更多的是对电网大数据的应用研究，往往容易忽视对电网大数据的管理，而正是对数字资产管理重要性的忽视，导致了上述问题的产生。只有解决了上述管理过程中存在的问题，才能更方便快速的进行电网大数据的分析应用，而集中式的管理方法显然无法实现这一目的。

## 2.2 基于区块链技术的解决方案

电网大数据体量大、种类多、数据源分布广泛、产生速度极快，继续使用传统集中式的数据管理方法效率低下且安全性较低，同时还会严重影响企业对电网大数据价值的挖掘。而区块链技术的产生，给电网大数据的数字资产管理提供了一条新的技术路径。通过采用分布式共识机制、链式区块结构、非对称加密算法等技术可实现区块链去中心化、去信任、信息可溯源以及信息难以篡改等特点，与电网大数据的数据特点以及管理需求十分吻合，故可基于区块链技术制定相应的解决方案来解决 2.1 节中所提到的问题。

### 2.2.1 区块链核心技术分析

（1）区块链的结构

通过链式结构将数据信息记录在区块链中，能够实现记录的连续性,如图 1 所示，区块链的基础构成是区块，每个区块通过区块头上的信息链接到前一个区块，形成链式结构[6]。其中区块是数据的一个集合，记录着一定时间内的每一条数据信息或交易内容。每个区块由两部分组成：1）区块头，记录着链接上一个区块的 hash 值，用于链接上一个区块，保证区块的连续性。同时，为了保证数据的可追溯性，还记录了每一个区块的时间戳；2）区块体，记录了规定时间内的所有数据信息和交易信息。区块链系统中的每个节点共同参与数据的管理和监督。

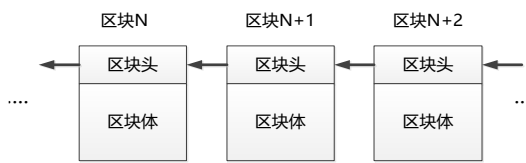

图 1 区块链结构图

（2）共识机制

与加密货币比特币不同，电网大数据属于电网企业内部数字资产，有严格的权限审查，故不能在公有链中对其进行管理。而联盟链网络的准入规则与电网数字资产管理理念吻合，所以本文拟使用联盟链的形式来构建用于管理电网大数据的区块链。在联盟链中使用较多的共识机制为股份授权证明（DPOS）共识机制，其基本原理是通过区块链系统中所有节点公平投票产生 101 个节点作为"受托人"负责轮流签署产生新的区块，并在后续的运营过程中可以根据各节点的表现情况决定是否重新进行投票[7]。

在电网企业中，由于可能存在不同省电力公司通过节点数量的优势影响投票结果，所以这种共识机制并不完全适用于电网机构。通过对 DPOS 共识机制进行改进，可得到符合电网实际情况的改进 DPOS 共识机制：在初始化节点时，对系统中所有节点进行考核评分，考核内容包括对电网数据的收集积极性、数据错漏率、电网运营情况等，根据评分对各节点进行排序，前 101 个节点作为首次签署区块的"受托人"，在此统称为数据记录节点。为了对产生的区块进行二次校验，确保数据的真实性、可信性，选取 101 名之后的前 20 个节点作为数据监督节点，剩下的节点则称为候选节点。完成对系统内节点的初始化之后，给所有节点设置一个信用积分，并赋予初始化。

在数字资产管理过程中，通过对节点的工作情况对信用积分进行更改，其具体方案将在后面的具体解决方案中介绍。定期通过对各节点所持有的信用积分进行排序，可实现新一轮的节点选取，避免了拉票行为的干扰。同时，将信用积分纳入各节点的年终考核指标中，直接与各节点的切身利益相关联，可以提高各节点参与管理和监督数字资产管理过程的积极性。

（3）非对称加密技术

区块链通过非对称加密技术来解决系统中各节点之间的信任问题。非对称加密算法会产生两个密钥：公开密钥和私有密钥[8]，每个节点均有属于自己的公钥和私钥，公钥会在全网中广播给其他节点，私钥只有节点自己拥有。如果用私钥对数据进行加密，则需要使用对应的公钥进行解密，同样的，若节点用公钥对数据进行加密，则需要使用对应的私钥进行解密。

在交易中的节点通过私钥对数据进行数字签名，其它节点可通过公钥解密来确认数据来源的真实性。

（4）分布式数据库

为了提高数据存储的可靠性，区块链技术采用分布式数据库技术，将每一份数据备份存储于分布在不同地区的存储单元中，这样，即使某一个存储单元遭到攻击或毁坏，也不会影响电网大数据的整体使用，且存储于被毁坏的存储单元中的数据可通过备份轻易恢复，大大提高了数据存储的可靠性。同时，分布式数据库分散了各节点对数据的调用请求，提高了数据库的并发性，降低了数据传输的成本，其高可扩展性也节省了大量的系统容量扩展成本。

### 2.2.2 解决方案制定

根据 2.1 节中对现存问题的总结与归纳，解决方案将从数字资产的记录、分享和安全这三个方面进行制定。

（1）数字资产的记录

电网大数据传统的数字资产管理方法在数据记录质量上出现问题的根本原因在于其层级式的管理架构，导致信息无法顺畅传递，信息在传递过程中容易失真。因此，要解决数字资产记录质量上的问题，首先需要打破这种层级式的管理架构，采用扁平的分布式管理架构。

通过区块链式存储结构和改进的 DPOS 共识机制使系统中数据记录过程受整个区块链系统中所有节点共同监督。数据的具体记录过程如下：

1）节点采集本地区电网中的电网数据，并以自己的公钥作为标识，向当值数据记录节点提交上传请求。

2）当值的数据记录节点对该节点的公钥进行验证，确认该节点具有上传数据的权限，并回复接收上传数据的请求。

3）节点用自己的私钥对数据的摘要进行数字签名，并用当值的数据记录节点的公钥对数据进行加密。

4）当值的数据记录节点通过自己的私钥解锁加密数据，并用数据上传节点的公钥解锁签名，将摘要与数据原文的 hash 值进行对比。确认数据由该节点上传，并将数据摘要以及数据上传节点的签名记录在区块中，并将加密数据存储于分布式数据库中。

5）每隔十分钟，当值的数据记录节点计算区块中数据记录的 Merkle 树以及 Merkle 根值[9-11]，并将自己的公钥注明在区块头上，然后将区块随机广播给当值的数据监督节点以及两个候选节点进行校验。

6）当值的数据监督节点和两个候选节点校验区块通过，向当值的数据记录节点发送认可信息。

7）当值的数据记录节点将新生成的区块链接到数据区块链中。

在区块中，每一条数据记录包含三个元素：公钥、数据摘要以及元数据。其中公钥用于确认数据上传节点身份以及访问权限。摘要为将数据进行 hash 计算得到的 hash 值，可用于校验数据的完整性，并且可作为在数据库中查找数据的索引。元数据部分存储着数据的描述性信息，例如数据的种类以及生成时间戳等，便于在区块链中按照类别查询数据，提高数据搜索速度。

通过信用积分奖惩制度可极大提高系统中各节点数据的上传和管理积极性，奖惩制度具体为：1）当节点上传的数据被查出不符合规定或存在数据造假的情况，将对其信用积分进行扣除，每上传一条错误数据扣一分，反之，若上传数据符合规定且无造假情况，则每条正确数据加一分；2）数据记录节点署名的区块若被数据监督节点或候选节点查出具有错误数据，则每校验出一个区块，扣除该数据记录节点一分，反之则加一分；数据监督节点和候选节点间相互监督，若校验结果相同，则每个节点加一分，若校验结果不同，以多数节点的校验结果为准，并对校验结果不同的节点扣除一分。由于信用积分将会影响之后节点的重选以及每年年终的考核，为了获得更高的信用积分，每个节点将会积极主动的执行系统所要求的正确的行为——记录优质的电网数据。因此，信用积分制度能够保证记录在数据库中的数据真实可信且质量达标。

（2）数字资产的分享

传统数字资产管理体系中，在节点间的数据分享过程往往由于信任和权限问题，需要耗费大量时间和资源进行权限审查和数据校验，严重影响电网数据价值挖掘工作。本方案采用非对称加密技术，解决了节点之间的信任问题，并确保了数据在传输过程中的安全性。

基于区块链的数据分享过程在节点之间直接进行，不需要第三方的介入，如图2所示。在交易中，数据发送方用接收方的公钥对数据进行加密产生数据的密文，确保除拥有私钥的接收方，其它节点均不可对加密数据进行解密，以此保证数据在传输过程中的安全性。同时将数据通过 hash 函数计算得到数据的摘要，并用自己的私钥对摘要进行数字签名。数据发送方将数据密文和数字签名一起发送给接收方，接收方收到后用发送方的公钥对数字签名解密得到数据摘要，验证发送方的身份。同时用自己的私钥解密数据密文得到原始数据，并通过 hash 函数计算出数据的摘要，通过对比两份摘要来快速验证数据的完整性。验证通过，则整个交易流程完成，交易信息按照数据记录的方法通过共识机制得到系统中所有节点的认可并记录在区块链中。

区块链中每条数据记录的公钥完成了数据资产的确权工作，每条数据由哪个节点上传，归属权为哪个节点，均可靠公钥信息进行验证，方便查找交易对象或者进行虚假数据问责。此外，由于区块链的去中心化及全网数据同步的特性，在不通过第三方中介的情况下，各节点均可对数字资产进行溯源，从每条数据的数据源，产生时间，到该数据经过哪些操作，均可在区块链中查询，交易与数据一旦被记录在链上，便不能被随意篡改，保证了数据的唯一性，遏制了在数据分享过程中的造假行为。

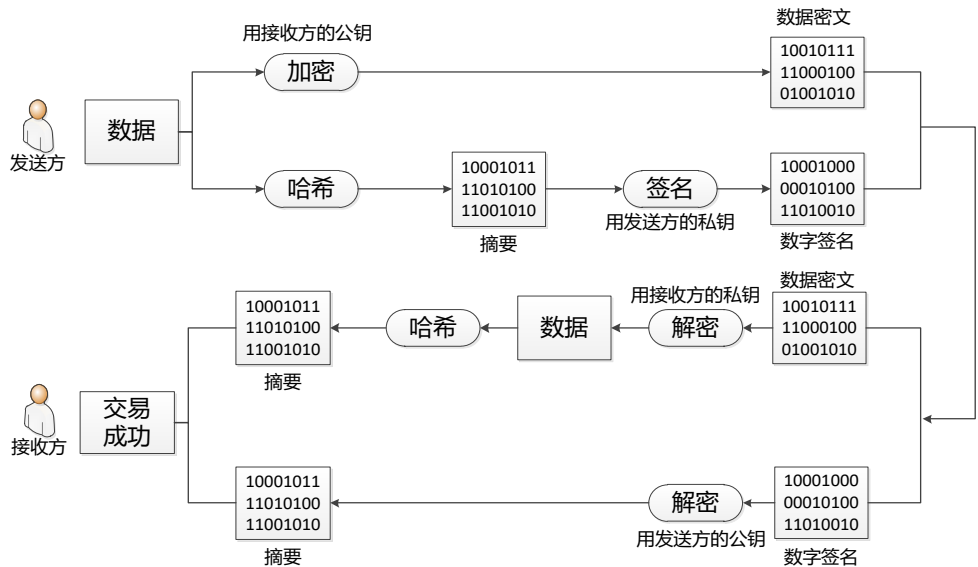

图2 数据分享流程图

（3）数字资产的安全

电网大数据的数字资产安全可分为数据存储安全以及数据传输安全，数据存储安全又分为数据在数据库中的安全以及数据在区块链上的记录安全。

通过非对称加密技术，将数据用数据所

有者的公钥加密存储于数据库中，即使数据库被黑客入侵，盗取了加密存储的数据密文，没有数据所有者独有的私钥便无法对密文进行解密，防止了数据的外泄。同时，由于每份数据均进行了备份，并存储于分布在不同区域的存储单元中，对单一或少数几个数据库的攻击或毁坏并不会影响整体数据的使用和恢复。另外，由于区块链的链式记录结构以及分布式的管理结构，一旦数据信息或交易信息被写入区块链中，任何对区块链中记录的更改都会导致更改点及之后的所有信息的改变，而系统中所有节点均有一份完整的区块链账本，能够十分方便的查验出记录的更改并修正，所以对单一或几个节点所拥有的账本的更改并不能篡改整个系统共同维护的记录，从而保证了数据记录的安全性。

而数据在传输过程中的安全性可参见数据分享的流程图，数据发送方用接收方的公钥对数据进行了加密，只有通过接收方独有的私钥才能解密，而在网络中进行传输的只有密文，并不会将原始数据在网络中进行传输，从而保证了数据在传输过程中的安全。

### 2.3 区块链化数字资产管理架构搭建

为了实现 2.2 节中指定的电网大数据数字资产管理方案，本文建立了电网大数据数字资产管理架构，如图 3 所示，该架构由分布式数据库、信用积分系统、数据区块链、数据记录节点、数据监督节点、候选节点以及第三方用户组成。在该管理架构的基础上，通过结合改进的 DPOS 共识机制以及信用积分体系，可实现对电网大数据整个生命周期创建、采集、组织、存储、利用以及清除各个环节的高效管理，同时保证了数据的安全和可靠。

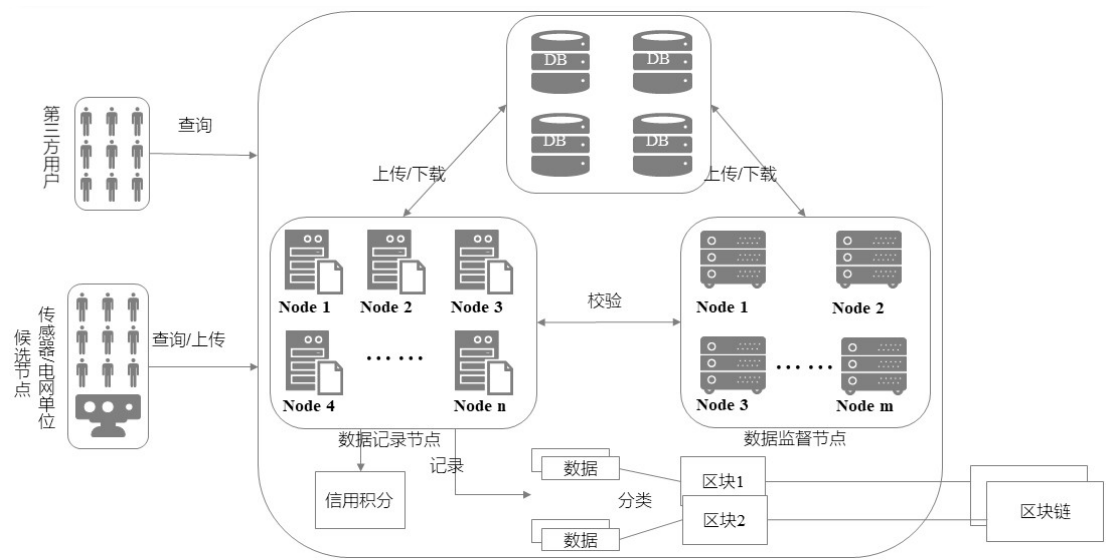

图 3 电网大数据数字资产管理架构

## 5 总结

电网大数据是支撑智能电网和电网企业安全稳定高效运行的基础，也是智能+时代电网企业拥有的丰富的数字资产。对电网大数据进行合理的数字资产管理，将促进电网企业对其应用的高效化。电网大数据数字

资产管理架构的建设必须融合最新的数据科学与新技术。区块链技术作为一个全网节点共同维护的分布式数据库，其去中心化、去信任、信息难以篡改等特性与电网大数据的数据特点和管理需求相吻合，在新的数字资产管理架构中具有极大的应用潜力。本文针对目前电网企业在电网大数据数字资产管理过程中出现的典型问题，设计了一种基于区块链技术的电网数字资产管理方法，和传统集中式的数字资产管理方法相比，该方法对电网大数据资产的管理更加合理高效，并使能了电网数字资产管理的众多功能。以区块链技术和数字资产管理理论为基础，其改变传统电网数据中心化管理模式，促进数据的分享，建立共识机制和信用积分激励机制，提升数据源产生和维护优质电网数据的积极性，优化电网数据存储模式，建立去中心化、共同维护、难以篡改的电网数据区块链。

本文所提出的区块链化的数字资产管理方法包括一个全新的分布式管理架构以及一套相对应的管理制度，旨在提高电网企业对数据管理的积极性和电网数据的质量。本文所提出的数字资产管理方法，只涉及数据记录、分享、安全等基本功能，可视为"电网数据资产管理区块链1.0"。该管理架构可作为对电网大数据的开发应用的基础架构，在此基础上可结合使用区块链智能合约技术和人工智能技术对电网数字资产架构进行进一步的更高层次开发，实现对数据的自动调取、分析、整合，在提升数字资产价值的同时，产生新的能够优化电网运营效率的"知识"，实现电网的数据智能与知识自动化。

# 参考文献

on the Theory and Applications of Cryptographic Techniques. Springer, Berlin, Heidelberg, 2004:541-554.

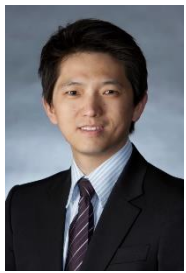
**张俊** 武汉大学电气工程学院教授.2003 年和 2005 年分别获得华中科技大学电气工程系学士与硕士学位.2008 年获得亚丽桑那州立大学电气工程博士学位. 主要研究方向为智能系统，人工智能，知识自动化，及其在智能电力和能源系统中的应用.本文通信作者.

E-mail: jun.zhang@qaii.ac.cn

(ZHANG Jun Professor at School of Electrical Engineering, Wuhan University. He received his B. E. and M. E. degrees in Electrical Engineering from Huazhong University of Science and Technology, Wuhan, China, in 2003 and 2005, respectively, and his Ph. D. in Electrical Engineering from Arizona State University, USA, in 2008. His research interest covers intelligent systems, artificial intelligence, knowledge automation, and their applications in intelligent power and energy systems. Corresponding author of this paper.)

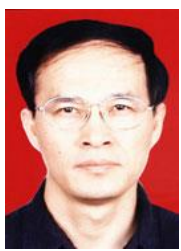
**王飞跃** 中国科学院自动化研究所复杂系统管理与控制国家重点实验室研究员.国防科技大学军事计算实验与平行系统技术研究中心主任. 主要研究方向为智能系统和复杂系统的建模、分析与控制.

E-mail: feiyue.wang@ia.ac.cn

(WANG Fei-Yue Professor at The State Key Laboratory of Management and Control for Complex Systems, Institute of Automation, Chinese Academy of Sciences. Director of the Research Center for Computational Experiments and Parallel Systems Technology, National University of Defense Technology. His research interest covers modeling, analysis, and control of intelligent systems and complex systems.)